\documentclass[prd,onecolumn,showpacs,floatfix,nofootinbib,amsmath,amssymb,floatfix]{revtex4}
\usepackage{graphicx,color,dcolumn,booktabs,bm,mathrsfs}
\usepackage{longtable,lscape}
\usepackage{txfonts}
\usepackage{overpic}
\usepackage{amssymb}
\usepackage{indentfirst}
\usepackage{feynmf}   %{feynmp}
\usepackage{slashed}  %for Feynman symbols
\usepackage{cases}
\usepackage{color}
\usepackage{multirow}
\usepackage{epstopdf}
\usepackage[colorlinks,
            citecolor=blue,
            anchorcolor=red,
            menucolor=red,
            linkcolor=red,
            filecolor=red,
            runcolor=red,
            urlcolor=blue,
            frenchlinks=red]{hyperref}
\usepackage{subfigure}

\usepackage{subfigure}

\begin{document}

\title{Revisit the $Z_c(4025)$ structure observed by BESIII in the $e^+e^- \to (D^*\bar{D}^*)^{\pm,0}\pi^{\mp,0}$ reactions at $\sqrt{s}=4.26$~GeV}

\author{Man-Yu Duan}
\affiliation{School of Physics and Microelectronics, Zhengzhou University, Zhengzhou, Henan 450001, China}
\affiliation{School of Physics, Southeast University, Nanjing 210094, China}

\author{Guan-Ying Wang}
\affiliation{School of Physics and Electronics, Henan University, Kaifeng 475004, China}

\author{En Wang}
\email{wangen@zzu.edu.cn}
\affiliation{School of Physics and Microelectronics, Zhengzhou University, Zhengzhou, Henan 450001, China}

\author{De-Min Li}\email{lidm@zzu.edu.cn}
\affiliation{School of Physics and Microelectronics, Zhengzhou University, Zhengzhou, Henan 450001, China}

\author{Dian-Yong Chen}
\affiliation{School of Physics, Southeast University, Nanjing 210094, China}

\begin{abstract}
Within the framework of the local hidden gauge formalism, we have reanalyzed the hidden-charm vector-vector meson $D^*\bar{D}^*$, $K^*\bar{K}^*$, and $\rho\rho$ interactions with the quantum numbers $I(J^P)=1(1^+)$. The estimation indicates that $D^*\bar{D}^*$ system could form a bound state around 4010~MeV. By comparing our theoretical estimation with the BESIII measurements of the $e^+e^- \to (D^*\bar{D}^*)^{\pm}\pi^{\mp}$ and $e^+e^- \to (D^*\bar{D}^*) ^{0}\pi^{0}$ reactions at $\sqrt{s}=4.26$~GeV , we find that the enhancement structure of the $Z_c(4025)$ in the $D^*\bar{D}^*$ invariant mass distribution observed by BESIII could be interpreted by a shallow $D^*\bar{D}^*$ bound state.
\end{abstract}

%\pacs{Valid PACS appear here}
% PACS, the Physics and Astronomy Classification Scheme.
% Valid PACS numbers may be entered using the \verb+\pacs{#1} command.

%\keywords{Baryons, Mesons, Resonances, Molecular states, Chiral unitary approach, Nonperturbative technique.}

\maketitle

%%%%%%%%%%%%%%%%%%%%%%%%%%%%%%%%%%%%%%%%%%%%%%%%%%%%%%%%%%%%%%%%%%%%%%%%%%%%%%%%%%%%%%%%%%%%%%%%%%%%%%%%%%%%%%%%%%%%%%%%%%%%%%%%%%%%%%%%%%%%%%%%%%%%%%%%%%%%%

\section{INTRODUCTION}
\label{sec:INTRODUCTION}

Since the observation of $X(3872)$ by the Belle Collaboration in 2003~\cite{Choi:2003ue}, a number of charmonium-like states, also named $XYZ$ states, have been discovered experimentally~\cite{Zyla:2020zbs}, the exotic properties of which have provided us an idea platform for understanding the non-perturbative property of the Quantum Chromodynamics (QCD)~\cite{Brambilla:2019esw,Chen:2016qju,Liu:2019zoy,Hosaka:2016pey}. For the nature of charmonium-like states, some exotic explanations have been proposed, such as tetraquark, molecular, hybrid, and kinematic effect, but the conventional charmonium explanations can not be discarded~\cite{Brambilla:2019esw,Chen:2016qju,Liu:2019zoy,Hosaka:2016pey,Brambilla:2010cs,Briceno:2015rlt,Guo:2017jvc}. The typical examples are $X(3915)$ and $X(3930)$ produced from the photon-photon fusion process~\cite{Belle:2009and,Belle:2005rte}, which can be assigned as the charmonia $\chi_{c0}(2P)$ and $\chi_{c2}(3P)$ states, respectively~\cite{Liu:2009fe}. But for a larger part of the charmonium-like states, both the exotic explanations and conventional charmonium assignments exist in the literatures, and their nature is still in doubt. Taking $X(4140)$ as an example, it was first observed by the CDF Collaboration in 2009~\cite{Aaltonen:2009tz} and has been interpreted as a conventional $c\bar{c}$ state~\cite{Hao:2019fjg,Chen:2016iua}, and the tetraquark state~\cite{Agaev:2017foq,Chen:2016oma,Wang:2018qpe,Wu:2016gas}, because of the large discrepancy of its measured widths from different experimental groups~\cite{LHCb:2016axx,Wang:2017mrt,Wang:2018djr}. However, the charmonium-like states with nozero isospin are particularly interesting since they are clear candidates for exotic states, which have attracted great experimental and theoretical concerns.

In 2014, the BESIII Collaboration observed a structure in the $\pi^\mp$ recoil mass spectrum of the process $e^+e^- \to (D^*\bar{D}^*)^{\pm}\pi^{\mp}$ at a center-of-mass energy of $4.26$~GeV, which was denoted as $Z_c^{\pm}(4025)$, with a mass of $4026.3 \pm  2.6 \pm 3.7$~MeV and a width of $24.8 \pm 5.6 \pm 7.7$~MeV~\cite{Ablikim:2013emm}. Later, its neutral partner, $Z_c(4025)^0$,  with a mass of $4025.5^{+2.0}_{-4.7} \pm 3.1$~MeV and a width of $23.0 \pm 6.0 \pm 1.0$~MeV , was observed in the $\pi^0$ recoil mass spectrum of the  process $e^+e^- \to (D^*\bar{D}^*)^{0}\pi^{0}$ at $\sqrt{s}=4.23$~GeV and $\sqrt{s}=4.26$~GeV by the BESIII Collaboration~\cite{Ablikim:2015vvn}. The observed mass of $Z_c(4025)$ are closed to the $(D^*\bar{D}^*)^\pm$ threshold, which indicates that the $Z_c(4025)$ could be interpreted as a deuteron-like molecular states~\cite{He:2013nwa,Chen:2015ata, Chen:2015jwa,Guo:2013sya}. Moreover, the isospin of $Z_c(4025)$ is one and the most possible constituent quark components are $c\bar{c} q\bar{q}$. Thus, the charmonium-like state $Z_c(4025)$ can be a good candidate of  tetraquark states ~\cite{Deng:2014gqa,Goerke:2016hxf,Wang:2013exa,Qiao:2013dda}. Besides these exotic states interpretation, the structure corresponding to $Z_c(4025)$ can also be reproduced by some kinematic mechanism, such as initial state pion emission mechanism~\cite{Wang:2013qwa, Chen:2013coa}, the reflection from the $P$-wave charmed meson $D_1(2420)$~\cite{Wang:2020axi}.

It is interesting to notice that many charmonium-like states have been observed around the thresholds of a pair of heavy hadrons, such as $X(3872)$ and $Z_c(3900)$ around $D\bar{D}^*$ threshold, $Z_{cs}(3985)/Z_{cs}(4000)$ around the $D_s\bar{D}^*/D_s^*\bar{D}$ threshold, and $X(3930)$ around $D_s\bar{D}_s$ threshold. Recently, we have shown that the existence of the $D\bar{D}$ bound state is supported by the measurements of the $e^+e^-\to J/\psi D\bar{D}$ and $\gamma\gamma \to D\bar{D}$~\cite{Wang:2019evy,Wang:2020elp}, and suggested to search for it in the $\Lambda_b\to \Lambda D\bar{D}$ reaction~\cite{Wei:2021usz}. As discussed in Ref.~\cite{Dong:2020hxe}, such structures should appear at any threshold where the interaction is attractive, thus the study of the near-threshold enhancement structure is crucial to deeply understand the interactions between heavy-hadrons, and further reveal the internal structures of the charmonium-like states.

As discussed in Ref.~\cite{Torres:2013lka}, enhancement structures close to the threshold of a pair of particles are sometimes identified as new particles, but they could also be due to  shallow molecular states below threshold. For instance, in Ref.~\cite{Ablikim:2009ac}, the BESIII Collaboration has observed a bump structure close to threshold in the $K^{*0}\bar{K}^{*0}$ mass distribution in the $J/\psi\to \eta K^{*0}\bar{K}^{*0}$ reaction, which can be interpreted as a signal of the $h_1$ resonance~\cite{Geng:2008gx,Xie:2013ula}. Similarly, in Ref.~\cite{Ablikim:2006dw}, an enhancement near threshold is observed in the $\omega\phi$ invariant mass spectrum from the doubly Okubo-Zweig-Iizuka (OZI) suppressed decays of $J/\psi\to\gamma\omega\phi$, which was interpreted as a signal of the $f_0(1710)$ resonance~\cite{Geng:2008gx,MartinezTorres:2012du}. In addition, an anomalous enhancement near the $\bar{p}\Lambda$ threshold was observed by BESIII Collaboration in the $\chi_{c0}\to\bar{p}\Lambda K^+$~\cite{Ablikim:2012ff}, which can be interpreted as a signal of the $K(1830)$ resonance~\cite{Wang:2020wap}.
 
Along this way, one could notice the structure $Z_c(4025)$ is near the threshold of $D^\ast \bar{D}^\ast$, thus  it may resulted from the $D^\ast \bar{D}^\ast$ shallow bound state below the threshold. In the framework of the local hidden gauge the vector-vector meson interactions have been studied ~\cite{Molina:2009ct}, and in the present work, we would like to generate dynamically the $D^*\bar{D}^*$ bound state with the quantum numbers of $I=1$ and $J^P=1^+$ by employing the formalism in Ref.~\cite{Molina:2009ct}. With this input, we will investigate the BESIII measurements of the $e^+e^- \to (D^*\bar{D}^*)^{\pm}\pi^{\mp}$ and $e^+e^- \to (D^*\bar{D}^*) ^{0}\pi^{0}$ reactions at $\sqrt{s}=4.26$~GeV by taking into account vector-vector meson interactions, which can deepen our understanding about the enhancement structure corresponding to $Z_c(4025)$, and the $D^*\bar{D}^*$ interaction.
 
This paper is organized as follows. In Sec.~\ref{sec:FORMALIISM}, we will show the formalism for the $e^+e^- \to (D^*\bar{D}^*)^{\pm}\pi^{\mp}$ and $e^+e^- \to (D^*\bar{D}^*)^{0}\pi^{0}$ reactions, and in Sec.~\ref{sec:RESULTS}, our numerical results and related discussions will be present. The last section will devoted to a short summary.

%%%%%%%%%%%%%%%%%%%%%%%%%%%%%%%%%%%%%%%%%%%%%%%%%%%%%%%%%%%%%%%%%%%%%%%%%%%%%%%%%%%%%%%%%%%%%%%%%%%%%%%%%%%%%%%%%%%%%%%%%%%%%%%%%%%%%%%%%%%%%%%%%%%%%%%%%%%%%

\section{FORMALISM}
\label{sec:FORMALIISM}

\subsection{Vector-vector Interaction}
We follow the approach of Refs.~\cite{Molina:2009ct,Aceti:2014kja}, which investigates the vector-vector interaction in the framework of the local hidden gauge formalism for hidden charm systems with quantum numbers $S=0$ and $C=0$. 
The Lagrangian is taken from the local hidden gauge formalism describing the interaction of vector mesons, 
\begin{equation}
\mathcal{L}=-\frac{1}{4}\langle V_{\mu\nu}V^{\mu\nu}\rangle\ ,
\label{eq:lvv}
\end{equation}
where the symbol $\langle ~ \rangle$ stands for the trace of $SU(4)$, and the tensor $V_{\mu\nu}$ is defined as
\begin{equation}
V_{\mu\nu}=\partial_{\mu}V_{\nu}-\partial_{\nu}V_{\mu}-ig[V_{\mu},V_{\nu}]\ 
\label{eq:vectensor}
\end{equation}
with $V_{\mu}$ to be 
\begin{equation}
V_\mu=\left(
\begin{array}{cccc}
\frac{\omega}{\sqrt{2}}+\frac{\rho^0}{\sqrt{2}} & \rho^+ & K^{*+}&\bar{D}^{*0}\\
\rho^- &\frac{\omega}{\sqrt{2}}-\frac{\rho^0}{\sqrt{2}} & K^{*0}&D^{*-}\\
K^{*-} & \bar{K}^{*0} &\phi&D^{*-}_s\\
D^{*0}&D^{*+}&D^{*+}_s&J/\psi
\end{array}
\right)_\mu\ .
\label{eq:vfields}
\end{equation}
The coupling constant $g$ will be given in following. 
Two different types of interactions can be derived from the Lagrangian of Eq.~\eqref{eq:lvv}, a contact interaction, coming from the $[V_{\mu},V_{\nu}]$ term,
\begin{equation}
\mathcal{L}^{\rm contact}=\frac{g^2}{2}\langle V_{\mu}V_{\nu}V^{\mu}V^{\nu}-V_{\nu}V_{\mu}V^{\mu}V^{\nu}\rangle\ ,
\label{eq:contact}
\end{equation}
and the three-vector vertex
\begin{equation}
\mathcal{L}^{(3V)}=ig\langle(\partial_{\mu}V_{\nu}-\partial_{\nu}V_{\mu})V^{\mu}V^{\nu} \rangle\ .
\label{eq:3v}
\end{equation}
The Lagrangian $\mathcal{L}^{(3V)}$ produces the $VV\rightarrow VV$ interaction by means of the exchange of one vector meson.

In the present work, we only take into account the channels coupled to the system with quantum numbers $I=1,J=1$, charm $C=0$ and strangeness $S=0$, which are $D^*\bar{D}^*$, $K^*\bar{K}^*$, $\rho\rho$, $\rho\omega$, $\rho J/\psi$, and $\rho\phi$. The potentials involving  the channels $\rho\omega$, $\rho J/\psi$, and $\rho\phi$ are exactly zero in this model~\cite{Molina:2009ct}. Thus, the potentials that will be used as the kernel to solve the Bethe-Salpeter (BS) equation, can be found in Table XIX of Ref.~\cite{Molina:2009ct} and Tables~IX and XXI of Ref.~\cite{Geng:2008gx}\footnote{These tables can be found in the arXiv version of Ref.~\cite{Geng:2008gx}.}.  For completeness, here we also present the explicit expressions of the potentials ($i=1,2,3$ correspond to the channels $D^*\bar{D}^*$, $K^*\bar{K}^*$,  and $\rho\rho$, respectively), 

\begin{gather}
V_{11}=3g_{D}^2+g_D^2\frac{\left[2M_{\omega}^2M_{\rho}^2+M_{J/\psi}^2(-M_{\omega}^2+M_{\rho}^2)\right]\left(4M_{D^*}^2-3s\right)}{4M_{J/\psi}^2 M_{\omega}^2 M_{\rho}^2}\ ,
\label{eq:t11} \\
V_{12}=\frac{9}{2}gg_{D}+gg_D\frac{2M_{D^*}^4+M_{D^*}^2\left(4M_{D_s^*}^2+2M_{K^*}^2-3s\right)+2M_{D_s^*}^2\left(2M_{\rho}^2-3s\right)}{4M_{D^*}^2M_{D_s^*}^2}\ ,
\label{eq:t12}\\
V_{13}=\frac{3}{\sqrt{2}}gg_{D}+gg_D\frac{2M_{D^*}^2+2M_{\rho}^2-3s}{2\sqrt{2}M_{D^*}^2}\ ,
\label{eq:t13}\\
V_{22}=3g^2+g^2\frac{\left[M_{\rho}^2M_{\phi}^2-M_{\omega}^2(M_{\phi}^2-2M_{\rho}^2)\right]\left(4M_{K^*}^2-3s\right)}{4M_{\phi}^2 M_{\omega}^2 M_{\rho}^2}\ ,
\label{eq:t22}\\
V_{23}=3\sqrt{2}g^2+g^2\frac{2M_{\rho}^2+2M_{K^*}^2-3s}{\sqrt{2}M_{K^*}^2}\ ,
\label{eq:t23}\\
V_{33}=6g^2+g^2\left(4-\frac{3s}{M_{\rho}^2}\right)\ ,
\label{eq:t33}
\end{gather} 
where $M_{\rho}$, $M_{\omega}$, $M_{\phi}$, $M_{K^*}$, $M_{D^*}$, $M_{D^*_s}$, and $M_{J/\psi}$ are the masses of the $\rho$, $\omega$, $\phi$, $K^*$, $D^*$, $D_s^*$, and $J/\psi$, respectively, and all these are taken from the Particle Data Group (PDG)~\cite{Tanabashi:2018oca}. $s$ is square of the center-of-mass energy of the coupled channel system. We take the constants $g=M_\rho/(2f_\pi)=4.17$ for light mesons,  and $g_D=M_{D^*}/(2f_D)=6.9$ with $f_D=206/\sqrt{2}=145.66$~MeV for $D^*$ meson, as used in Ref.~\cite{Molina:2009ct}.
With the rules of heavy quark spin symmetry~\cite{Wise:1992hn}, which can be obtained from the impulse approximation at the quark level assuming the $s$ and $c$ as spectators, the $D^*\bar{D}^*\to D^* \bar{D}^*$ is mediated by $J/\psi$ exchange $(c\bar{c})$ in analogy to the $\phi$ exchange in $K^*\bar{K}^*\to K^*\bar{K}^*$~\cite{Liang:2014eba}. Thus, the vector exchange term in Eq.~(\ref{eq:t11}) can be replaced by $g_D^2 m^2_{D^*}/m^2_{J/\psi}$  as used in Ref.~\cite{Aceti:2014kja}. And we use here the normal $g$ coupling in the contact and exchange terms of the other potentials for simplicity.

With the potentials in Eqs.~(\ref{eq:t11})-(\ref{eq:t33}), one can solve the Bethe-Salpeter equation in coupled channels, which is
\begin{equation}
T=[1-VG]^{-1}V\ ,
\label{eq:BS}
\end{equation}
where $V$ is a $3\times3$ matrix. The matrix $G$ is the $3\times 3$ diagonal matrix whose elements are the two-meson loop function given by
\begin{equation}
G_i=i\int\frac{d^4q}{(2\pi)^4}\frac{1}{q^2-m_1^2+i\epsilon}\frac{1}{(q-P)^2-m_2^2+i\epsilon}\ ,
\label{eq:loopex}
\end{equation}
with $m_1$ and $m_2$ the masses of the two mesons involved in the loop in the $i$-th channel. $q$ is the four-momentum of the meson, and $P$ is the total four-momentum of the meson-meson system.  In this work, we use the dimensional regularization method as Refs.~\cite{Aceti:2014kja,Duan:2020vye,Molina:2009ct}, and the function $G_i$ can be rewritten  as,
\begin{eqnarray}
G_i&=&\frac{1}{16\pi^2}\left[\alpha_i+\log\frac{m_1^2}{\mu^2}+\frac{m_2^2-m_1^2+s}{2s}\log\frac{m_2^2}{m_1^2}\right. \nonumber\\
&&+\frac{|\vec{p}\,|}{\sqrt{s}}\left(\log\frac{s-m_2^2+m_1^2+2|\vec{p}\,|\sqrt{s}}{-s+m_2^2-m_1^2+2|\vec{p}\,|\sqrt{s}}\right. \nonumber \\
&&+\left. \left. \log\frac{s+m_2^2-m_1^2+2|\vec{p}\,|\sqrt{s}}{-s-m_2^2+m_1^2+2|\vec{p}\,|\sqrt{s}}\right)\right] ,
\label{eq:loopexdm}
\end{eqnarray}
where $\vec{p}\,$ is the three-momentum of the meson in the centre of mass frame, and
\begin{equation}
|\vec{p}\,|=\frac{\sqrt{\left[s-(m_1+m_2)^2\right]\left[s-(m_1-m_2)^2\right]}}{2\sqrt{s}} .
\end{equation}

\begin{figure}[h]
  \centering
\includegraphics[scale=0.85]{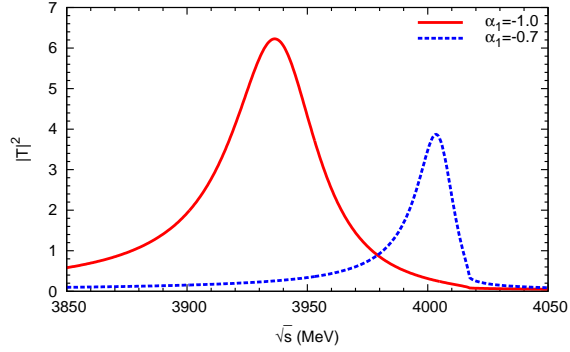}
   \caption{(Color online) The modulus squared of the amplitude $|t_{D^*\bar{D}^* \to D^*\bar{D}^*}|^2$ calculated with different $\alpha_{1}$. the red solid and blue dashed curves correspond to $\alpha_1=-1.0$ and $\alpha_1=-0.7$, respectively.}
    \label{fig:T}
  \end{figure}

Here we fix $\mu=1000$~MeV for all the channels, and take the subtraction constants for the loops containing SU(3) mesons $\alpha_2=\alpha_3=-1.65$ in order to find the position of the pole $f_2(1275)$, as used in Refs.~\cite{Geng:2008gx,Molina:2009ct}. We take the subtraction constant $\alpha_1$ for heavy channel as a free parameter. In Fig.~\ref{fig:T}, we show the modulus squared of the amplitudes with different values of  $\alpha_1$, and one can find a clear peak of the $I=1$ $D^*\bar{D}^*$ bound state.

In the complex plane, for a general $\sqrt{s}$, the loop function in the second Riemann sheet can be written as~\cite{Roca:2005nm},
\begin{equation}
\label{eq:lloopexdm}
G_i^{II}(\sqrt{s})=G_i^{I}(\sqrt{s})+i\frac{|\vec{p}\,|}{4\pi\sqrt{s}},\qquad {\rm Im}(|\vec{p}\,|)>0,
\end{equation}
where $G_i^{II}$ refers to the loop function in the second Riemann sheet, and $G_i^{I}$ is the one in the first Riemann sheet as given by Eq.~\eqref{eq:loopexdm} for the $i$-th channel. When searching for the poles, we use $G_i^{I}$ for Re$(\sqrt{s})<m_1+m_2$, and use $G_i^{II}$ for Re$(\sqrt{s})>m_1+m_2$. When the pole is not very far from the real axis, it occurs in $\sqrt{s_{p}}=(M\pm i \Gamma/2)$. The meaningful physical quantity is the value of the amplitudes for real $\sqrt{s}$, only poles not too far from the real axes would be easily identified experimentally as a resonance. In this case, the amplitude in Eq.~\eqref{eq:BS} close to a pole looks like
\begin{equation}
T_{ij}\approx \frac{g_i g_j}{s-s_{p}}\ , 
\label{eq:polegg}
\end{equation}
where the coupling constants $g_i$ for the resonance to $i$-th channel are calculated by means of the residues of
the amplitudes. We calculate the pole position and the couplings $g_i$ for different $\alpha_1$, which are listed in Table~\ref{tab:res}.

 \begin{table}[h]
   \begin{center}
  \caption{The pole positions and couplings $g_{i}$ in units of MeV for different $\alpha_1$, the number 1, 2, and 3 stand for channels $D^*\bar{D}^*$, $K^*\bar{K}^*$, and $\rho\rho$, respectively.}
  \label{tab:res}  %\centerline{$\sqrt{s}_{pole}=4010.55 + i 7.38$}
  \vspace{0.2cm}
  \begin{tabular}{ccccc}
 \toprule[1 pt]
  $\alpha_1$ &\qquad $\sqrt{s_p}$ &\qquad $g_1$ &\qquad $g_2$ &\qquad $g_3$\\
 \midrule[1 pt] 
  -1.0   &\qquad   3938.25 + $i$21.53   &\qquad   20702.31 - $i$825.13   &\qquad   2817.57 - $i$181.82   &\qquad   1354.20 + $i$108.94\\
  -0.7   &\qquad   4005.33 + $i$9.53   &\qquad   12467.81 - $i$2188.74   &\qquad   1877.11 - $i$462.52   &\qquad   926.81 - $i$95.52\\
  -0.65   &\qquad   4010.55 + $i$7.38   &\qquad   10964.94 - $i$2438.74   &\qquad   1655.87 - $i$497.48   &\qquad   823.45 - $i$125.99\\
  \bottomrule[1 pt]
  \end{tabular}
  \end{center}
  \end{table}

\subsection{The mechanism of the process $e^+e^-\to D^*\bar{D}^* \pi$}
\label{sec:process}

In the production of a $D^*\bar{D}^*$ bound state, the amplitude depends on the invariant mass of $D^*\bar{D}^*$\cite{Torres:2013lka}.  In particular, for the $e^+e^- \to (D^*\bar{D}^*)^{\pm}\pi^{\mp}$ reaction, the differential cross section depending on $D^*\bar{D}^*$ invariant mass is given by~\cite{Gamermann:2007mu},
\begin{align}\label{eq:sigma}
\frac{d\sigma}{d M_{D^*\bar{D}^*}} = V_p\frac{m^2_e}{s\sqrt{s}}k\tilde{q}\left|t_{D^*\bar{D}^* \to D^*\bar{D}^*}\right|^2,
\end{align}
where $V_p$ is a normalization factor, $m_e$ is the mass of electron~\cite{Tanabashi:2018oca}, $\sqrt{s}=4.26$~GeV. The momentum $k$ and $\tilde{q}$ are the pion momentum in the $e^+ e^-$ CM frame and the $D^*$ momentum in the $D^*\bar{D}^*$ CM frame, which are 
\begin{align}
k&=\frac{\lambda^{1/2}(s,m^2_\pi,M^2_{D^*\bar{D}^*})}{2\sqrt{s}},\\
\tilde{q}&=\frac{\lambda^{1/2}(M^2_{D^*\bar{D}^*},m^2_{D^*},m^2_{\bar D^*})}{2M_{D^*\bar{D}^*}},
\end{align}
respectively. The amplitude $t_{D^*\bar{D}^* \to D^*\bar{D}^*}$ can be obtained by solving Eq.~\eqref{eq:BS}\footnote{It should be stressed that the transition amplitude $t_{D^*\bar{D}^* \to D^*\bar{D}^*}$ obtained by Eq.~\eqref{eq:BS} depends on the $D^*\bar{D}^*$ invariant mass $M_{D^*\bar{D}^*}$, and the square of the vector-vector channels $s$ appeared in SubSect~\ref{sec:FORMALIISM} should be replaced by $M^2_{D^*\bar{D}^*}$.}. For the neutral channel $e^+e^- \to (D^*\bar{D}^*)^{0}\pi^{0}$ reaction, we can use another normalization factor $V_p^{\prime}$ in place of $V_p$ of Eq.~\eqref{eq:sigma}.

%%%%%%%%%%%%%%%%%%%%%%%%%%%%%%%%%%%%%%%%%%%%%%%%%%%%%%%%%%%%%%%%%%%%%%%%%%%%%%%%%%%%%%%%%%%%%%%%%%%%%%%%%%%%%%%%%%%%%%%%%%%%%%%%%%%%%%%%%%%%%%%%%%%%%%%%%%%%%

\section{RESULTS AND DISCUSSIONS}
\label{sec:RESULTS}

  \begin{figure}[h]
  \centering
  \subfigure{\includegraphics[scale=0.85]{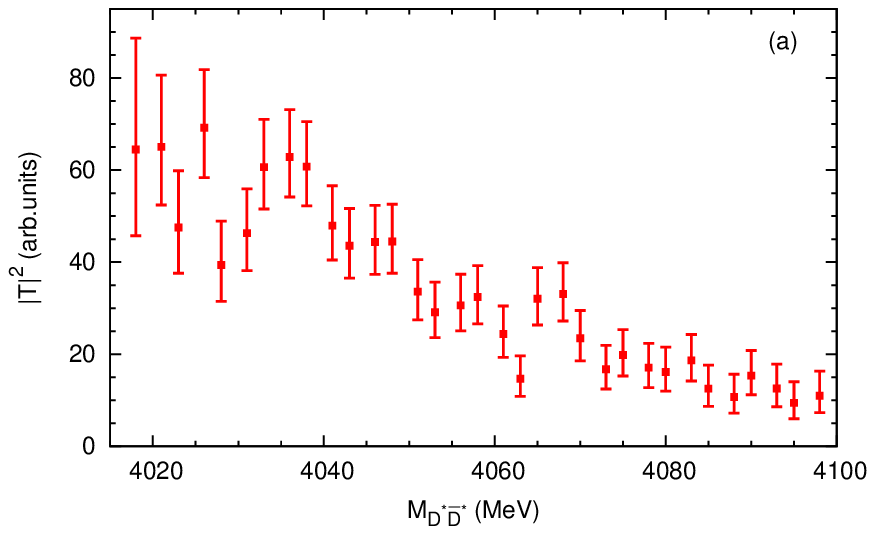}\label{fig:phsp+}}
  \subfigure{\includegraphics[scale=0.85]{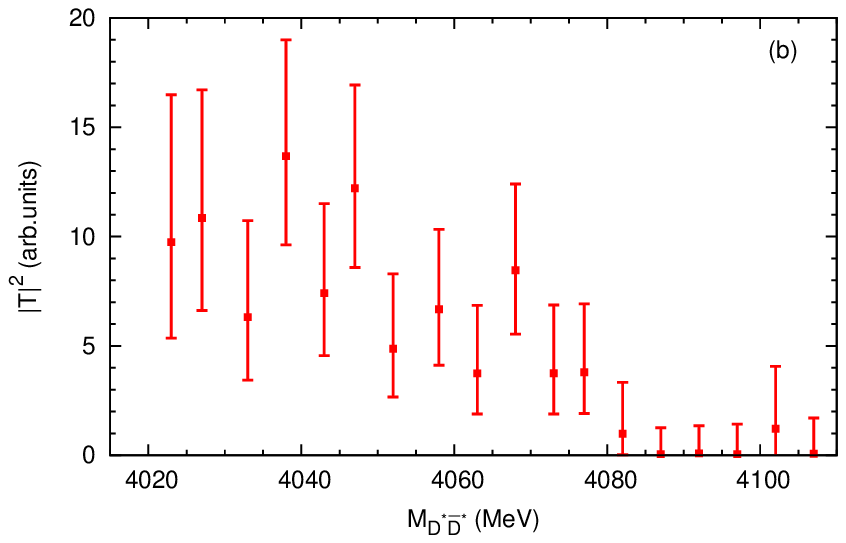}\label{fig:phsp0}}
  \caption{\small{The $D^*\bar{D}^*$ invariant mass distributions of the $e^+e^- \to (D^*\bar{D}^*)^{\pm,0}\pi^{\mp,0}$ reactions measured by BESIII divided by phase space factor ${k\tilde{q}}/{s\sqrt{s}}$ of Eq.~\eqref{eq:sigma}. (a) the $e^+e^- \to (D^*\bar{D}^*)^{\pm}\pi^{\mp}$ reaction~\cite{Ablikim:2013emm},  and (b) the $e^+e^- \to (D^*\bar{D}^*)^{0}\pi^{0}$ reaction~\cite{,Ablikim:2015vvn}.}}
  \label{fig:phsp}
  \end{figure}
  
For the first step, we divide the measured data of the $D^*\bar{D}^*$ invariant mass distributions from the BESIII Collaboration by a phase space factor ${k\tilde{q}}/{s\sqrt{s}}$, which are present in Fig.~\ref{fig:phsp}. The left and right panels correspond to the $D^*\bar{D}^*$ invariant mass distributions of the reactions $e^+e^- \to (D^*\bar{D}^*)^{\pm}\pi^{\mp}$ and $e^+e^- \to (D^*\bar{D}^*)^{0}\pi^{0}$, respectively, while the experimental data are taken from Refs.~\cite{Ablikim:2013emm,Ablikim:2015vvn}. One can find that there is no significant peak structure around 4025~MeV, and both distributions peak at the threshold of 4013.7~MeV for $D^{*0}\bar{D}^{*0}$ and 4020.5~MeV for $D^{*+}D^{*-}$, which implies that the shallow bound state below the threshold may play an important role for the enhancement structure of $Z_c(4025)$.

  For the second step, we perform a direct comparison of our results with the events measured by BESIII Collaboration. Here we take the two different normalization factors $V_p$ and $V_p^{\prime}$ for the reactions $e^+e^- \to (D^*\bar{D}^*)^{\pm}\pi^{\mp}$ and $e^+e^- \to (D^*\bar{D}^*)^{0}\pi^{0}$, respectively. In our model, there is another free parameter, the $\alpha_1$ in Eq.~\eqref{eq:loopexdm}. We will fit our model to the experimental data of BESIII Collaboration~\cite{Ablikim:2013emm,Ablikim:2015vvn} in the following\footnote{We do not consider the data points below the $D^*\bar{D}^*$ threshold in our fit. In addition, the last six data points of the  $e^+e^- \to (D^*\bar{D}^*)^{0}\pi^{0}$ reaction are also not considered, since their values are zero due to the detection efficiency.}.

\begin{figure}[h]
  \centering
  \subfigure{\includegraphics[scale=0.85]{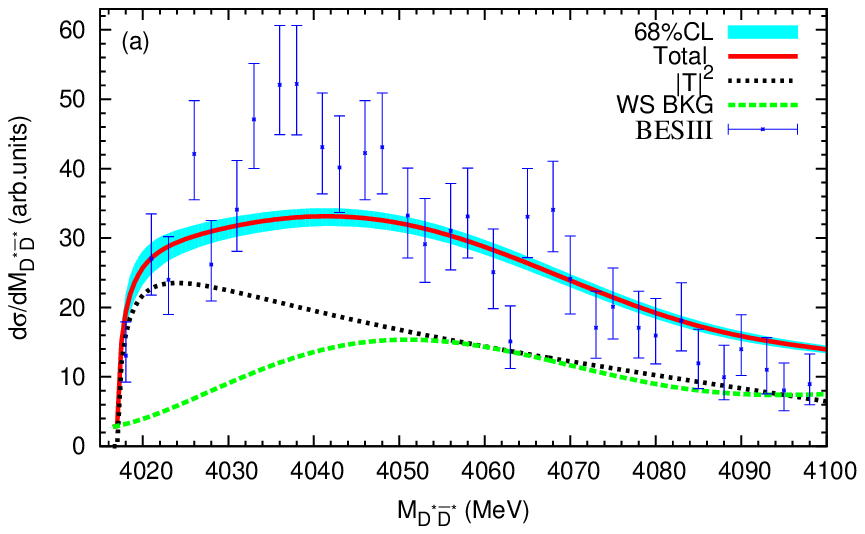}\label{fig:sigma+}}
  \subfigure{\includegraphics[scale=0.85]{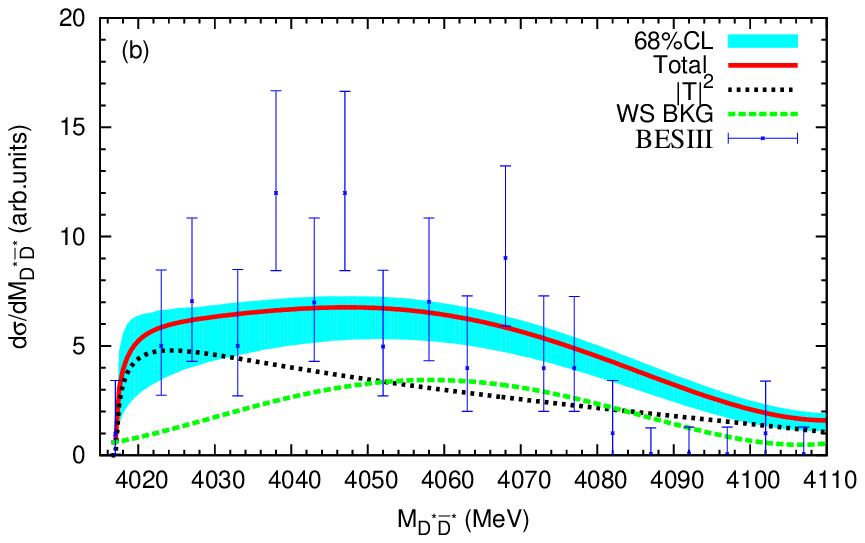}\label{fig:sigma0}}
  \caption{\small{The $D^*\bar{D}^*$ invariant mass distributions with the fitted parameters $V_p$/$V_p^\prime$ and $\alpha_1$ of Table~\ref{tab:fit}. The curves labeled as `Total' show the calculated results, The `$|T|^2$' curves stand for the contribution from the shallow $D^*\bar{D}^*$ bound state, and the `WS BKG' curves are the events of the wrong sing background taken from Refs.~\cite{Ablikim:2013emm,Ablikim:2015vvn}. The experimental data labeled as `BESIII' show the BESIII measurements~\cite{Ablikim:2013emm,Ablikim:2015vvn}, and the error bands labeled as `68\%CL' correspond to the 68\%   confidence level of our results due to the uncertainties of the fitted parameters. Diagrams (a) and (b) correspond to the $e^+e^- \to (D^*\bar{D}^*)^{\pm}\pi^{\mp}$ and the $e^+e^- \to (D^*\bar{D}^*)^{0}\pi^{0}$ reactions, respectively. }}
  \label{fig:sigma}
  \end{figure}

It should be pointed out that the components of the wrong sign background events are also taken into account in out fit, as pointed  by the BESIII Collaboration~\cite{Ablikim:2013emm,Ablikim:2015vvn}. The $\chi^2/d.o.f.$ are $1.74$ and $0.55$ for the reactions $e^+e^- \to (D^*\bar{D}^*)^{\pm}\pi^{\mp}$ and $e^+e^- \to (D^*\bar{D}^*)^{0}\pi^{0}$, respectively. The fitted parameters are tabulated in Table~\ref{tab:fit} (Fit A-charge, Fit A-neutral), and both fits favor the value of $a_\mu =-0.65$, which corresponds to a $D^*\bar{D}^*$ bound state with a mass of 4010.6~MeV and a width of 14.7~MeV, as shown in Table~\ref{tab:res}. With the fitted parameters, we have calculated the $D^*\bar{D}^*$ invariant mass distributions for the $e^+e^- \to (D^*\bar{D}^*)^{\pm}\pi^{\mp}$ and  $e^+e^- \to (D^*\bar{D}^*)^{0}\pi^{0}$ reactions, as shown in Fig.~\ref{fig:sigma}, and for comparison we also plot the BESIII measurements~\cite{Ablikim:2013emm,Ablikim:2015vvn}. One can see that our results are in reasonable agreement with the BESIII data considering the experimental uncertainties, which implies that the $D^*\bar{D}^*$ bound state, dynamically generated from the vector-vector meson interactions, plays an important role for the enhancement structure near the threshold. 

  \begin{figure}[h]
  \centering
  \subfigure{\includegraphics[scale=0.85]{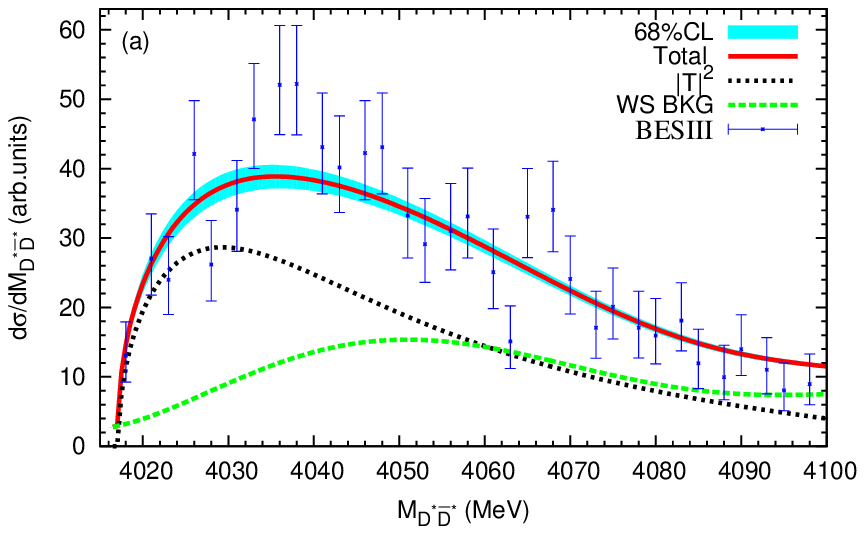}\label{fig:BW-sigma+}}
  \subfigure{\includegraphics[scale=0.85]{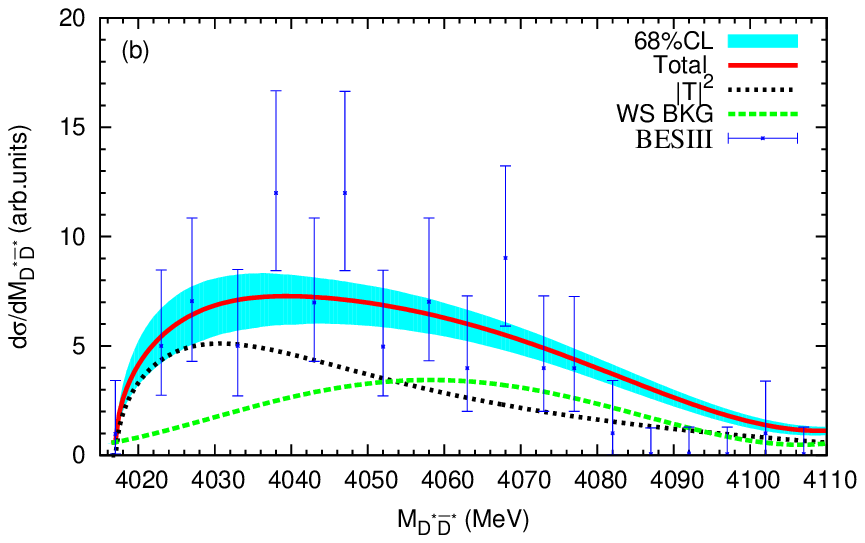}\label{fig:BW-sigma0}}
  \caption{\small{The $D^*\bar{D}^*$ invariant mass distributions using Eq.~\eqref{eq:polegg}, with the fitted parameters of Table~\ref{tab:fit}. The explanations of the curves are the same as the ones of Fig.~\ref{fig:sigma}.}}
  \label{fig:BW-sigma}
  \end{figure}

\begin{table}[h]
    \caption{The model parameters obtained by fitting to the BESIII measurements~\cite{Ablikim:2013emm,Ablikim:2015vvn}.}
  \label{tab:fit}
  \begin{center}
  \vspace{0.2cm}
  \centering
  \begin{tabular}{p{2.2cm}<\centering p{2.5cm}<\centering p{3.0cm}<\centering p{2.5cm}<\centering p{1.5cm}<\centering }
  \toprule[1pt]
 &$\alpha_1$ &$V_p$ or $V_p^\prime$ &$\Gamma_0$  & $\chi^2/d.o.f$\\
 \midrule[1pt]
 Fit A-charge   &   $-0.644\pm 0.032$   &   $1344.5\pm 122.6$   & -  & 1.74\\ 
  Fit A-neutral   &   $-0.647\pm 0.020$   &   $272.0\pm 121.0$   & - & 0.55\\
     Fit B-charge   &   -   &   $1363.5\pm 96.3$   &  $27.4\pm 0.9$&  1.03\\
    Fit B-neutral   &   -   &   $275.7\pm 98.3$   & $30.4\pm 7.3$ & 0.50\\
  \bottomrule[1pt]
  \end{tabular}
  \end{center}

  \end{table} 

It should be stressed that, in addition to the channels $K^*\bar{K}^*$ and $\rho\rho$, there should be other light mesons channels, which could enlarge the width of this bound state.  Then we repeat the fits by considering the width of the $D^*\bar{D}^*$ bound state as free parameter $\Gamma_0$, and replacing the amplitude $t_{D^*\bar{D}^* \to D^*\bar{D}^*}$ of Eq.~(\ref{eq:sigma}) by Eq.~(\ref{eq:polegg}), with $\sqrt{s_p}=4010.6+i\Gamma_0/2$ and the couplings of Table~\ref{tab:res}. The fit parameters are tabulated in Table~\ref{tab:fit} (Fit B-charge, Fit B-neutral), and one can found the width is about $27\sim 30$~MeV. With the parameters, we calculate the $D^*\bar{D}^*$ mass distributions shown in Fig.~\ref{fig:BW-sigma}
 and one can find that our results are in good agreement with the BESIII data as shown in Fig.~\ref{fig:BW-sigma}, which implies that the shallow $D^*\bar{D}^*$ bound state can provide an excellent interpretation for the enhancement structures of the $Z_c(4025)$.

%%%%%%%%%%%%%%%%%%%%%%%%%%%%%%%%%%%%%%%%%%%%%%%%%%%%%%%%%%%%%%%%%%%%%%%%%%%%%%%%%%%%%%%%%%%%%%%%%%%%%%%%%%%%%%%%%%%%%%%%%%%%%%%%%%%%%%%%%%%%%%%%%%%%%%%%%%%%%

\section{Conclusions}
\label{sec:conc}

In the present work, we have analyzed the  $e^+e^- \to (D^*\bar{D}^*)^{\pm}\pi^{\mp}$ and $e^+e^- \to (D^*\bar{D}^*)^{0}\pi^{0}$ reactions at $\sqrt{s}=4.26$~GeV reported by the BESIII Collaboration by taking into account the hidden-charm vector-vector meson interaction of quantum numbers $I(J^P)=1(1^+)$ within the framework of the local hidden gauge formalism.  We have shown that the $D^*\bar{D}^*$ invariant mass distributions of the BESIII measurements divided by the phase space have no significant peak around 4025~MeV, while both distributions peak at the threshold of 4013.7~MeV for $D^{*0}\bar{D}^{*0}$ and 4020.5~MeV for $D^{*+}D^{*-}$, which implies that one shallow state below the $D^*\bar{D}^*$ threshold may contribute to the enhancement structures of $Z_c(4025)$.

Employing the transition amplitude obtained by solving the Bethe-Salpeter equation with the potentials of the local hidden gauge formalism, we fit to the BESIII measurements of the $e^+e^- \to (D^*\bar{D}^*)^{\pm}\pi^{\mp}$ and $e^+e^- \to (D^*\bar{D}^*)^{0}\pi^{0}$ reactions. Our calculations are in reasonable agreement with the BESIII data by considering the experimental uncertainties. One $D^*\bar{D}^*$ bound state with a mass of 4010.6~MeV and a width of $14.7$~MeV,  dynamically generated from the hidden-charm vector-vector meson interaction could play an important role for the enhancement structures of $Z_c(4025)$.

Since there are other decay channels, in addition to the vector-vector channels considered here, for the $D^*\bar{D}^*$ bound state, its width should be larger than the one obtained by the present coupled channel Bathe-Salpeter equation. Taking into account this issue, we repeat the fits by regarding the width as a free parameter, and using the obtained pole position and the coupling constants. We can obtain an excellent description for the BESIII measurements, which implies that a shallow $D^*\bar{D}^*$ bound state with a mass of 4010.6~MeV and a width of  $27\sim 30$~MeV, could be used to interpret the enhancement structures near the threshold.

%%%%%%%%%%%%%%%%%%%%%%%%%%%%%%%%%%%%%%%%%%%%%%%%%%%%%%%%%%%%%%%%%%%%%%%%%%%%%%%%%%%%%%%%%%%%%%%%%%%%%%%%%%%%%%%%%%%%%%%%%%%%%%%%%%%%%%%%%%%%%%%%%%%%%%%%%%%%%

\begin{acknowledgments}
We would like to acknowledge the fruitful discussions with Eulogio Oset, Jun-Xu Lu, Chu-Wen Xiao, and Ju-Jun Xie. This work is partly supported by the National Natural Science Foundation of China under Grants Nos. 11775050, 11505158.  It is also supported by the Key Research Projects of Henan Higher Education Institutions under No. 20A140027, the Project of Youth Backbone Teachers of Colleges and Universities of Henan Province (2020GGJS017), the Natural Science Foundation of Henan (212300410123), the Youth Talent Support Project of Henan (2021HYTP002), the Fundamental Research Cultivation Fund for Young Teachers of Zhengzhou University (JC202041042), and the Academic Improvement Project of Zhengzhou University.
\end{acknowledgments}

\end{document}